\documentclass[letterpaper]{JHEP3}
\usepackage{amssymb,amsfonts}
\usepackage{cite}
\usepackage{epsfig}

\newcommand{\R}{{\bf R}}

\newcommand{\CF}{{\cal F}}

\newcommand{\CM}{{\cal M}}

\newcommand{\CV}{{\cal V}}

\newcommand{\bk}{{\bf k}}
\newcommand{\bx}{{\bf x}}
\newcommand{\bw}{{\bf w}}

\newcommand{\p}{\partial}
\newcommand{\newton}{G_{\!N}^{}}
\newcommand{\dif}{{\rm Diff}_{\!\CF}^{}}

\renewcommand{\tilde}[1]{\widetilde{#1}}
\newcommand{\be}{\begin{equation}}
\newcommand{\ee}{\end{equation}}
\newcommand{\bea}{\begin{eqnarray}}
\newcommand{\eea}{\end{eqnarray}}
\usepackage{graphicx}
\usepackage{latexsym}

\newcommand{\ie}{{\it i.e.}}
\newcommand{\eg}{{\it e.g.}}
\title{Spectral Dimension of the Universe\\ 
in Quantum Gravity at a Lifshitz Point}
\author{Petr Ho\v{r}ava\\
Berkeley Center for Theoretical Physics and Department of Physics\\
University of California, Berkeley, CA, 94720-7300\\
and\\
Theoretical Physics Group, Lawrence Berkeley National Laboratory\\
Berkeley, CA 94720-8162, USA}
\abstract{We extend the definition of ``spectral dimension'' (usually 
defined for fractal and lattice geometries) to theories on smooth spacetimes 
with anisotropic scaling.  We show that in quantum gravity dominated by a 
Lifshitz point with dynamical critical exponent $z$ in $D+1$ spacetime 
dimensions, the spectral dimension of spacetime is equal to 
$$
d_s=1+\frac{D}{z}.
$$
In the case of gravity in $3+1$ dimensions presented in arXiv:0901.3775, which 
is dominated by $z=3$ in the UV and flows to $z=1$ in the IR, the spectral 
dimension of spacetime flows from $d_s=4$ at large scales, to $d_s=2$ at 
short distances.  Remarkably, this is the qualitative behavior of $d_s$ found 
numerically by Ambj\o rn, Jurkiewicz and Loll in their {\it causal dynamical 
triangulations\/} approach to quantum gravity.}  
\begin{document}
\section{Introduction}

The idea that the effective spacetime dimension can change with the scale 
is not new.  

One simple thing that can happen as we probe spacetime at shorter distances is 
that extra dimensions can emerge.  The fact that our macroscopic Universe 
appears, to a good appproximation, four-dimensional is then viewed as a result 
of course graining.  Such extra dimensions can be of the Kaluza-Klein type 
\cite{kk1,kk2,kk3}, or our observed universe can be the boundary of a 
higher-dimensional space \cite{hw1,hw2,rs1,rs2}, or a higher codimension 
brane, perhaps with additional warping of the full geometry.  

Another intriguing possibility is that the nature of the four macroscopic 
spacetime dimensions themselves may qualitatively change with the changing 
scale.  The poor short-distance behavior of general relativity has often been 
interpreted as an indication that something radical must happen to spacetime 
at short distances.  It has been speculated that at some characteristic scale 
(often related to the Planck scale), the smooth geometry of spacetime could be 
replaced by a discrete structure, or exhibit some form of fractal behavior, or 
a stringy generalization of geometry, or that the short-distance nature of 
spacetime might be non-geometric altogether.  This 
picture is futher supported by our current understanding of string theory, in 
which the macroscopic spacetime -- or at least space -- can often be viewed as 
an emergent concept.  

In recent numerical simulations of lattice quantum gravity in the framework 
of causal dynamical triangulations (CDT) \cite{ajlem,ajlprin,ajlrec}, an 
interesting phenomenon has been observed:  The system exhibits a phase in 
which the effective spacetime dimension is four at large scales, but changes 
continuously to two at short distances \cite{ajl}.   The four-dimensional 
nature of spacetime at large scales indicates that the model does exhibit a 
good long-distance continuum limit.  However, the interpretation of the 
effective change in dimension at shorter scales is not clear.  Perhaps the 
geometry undergoes a dynamical dimensional reduction, or develops a foamy 
structure at short distances.  The lattice methods of dynamical triangulations 
do not offer enough analytical control over the details of the dynamics of 
geometry, and it would be desirable to compare the findings against an 
analytical tool in a continuum framework.  

The CDT approach to quantum gravity has one distinguishing feature:  The 
triangulations are restricted to conform to a preferred causal structure, 
given by a preferred foliation by slices of constant time.  This restriction 
is motivated by the desire to maintain causality and leads to the suppression 
of baby universes, which -- when present -- are believed to be responsible for 
the pathological branched-polymer scaling in the continuum limit.  

The preferred causal structure imposed on spacetimes in the CDT framework is 
quite reminiscent of the symmetries in the recently proposed Lifshitz phase of 
quantum gravity \cite{mqc,lif}.  This theory -- defined in the path-integral 
framework -- exhibits an anisotropic scaling of spacetime at short distances.  
The degree of anisotropy is measured by the dynamical critical exponent $z$, 
which changes from $z=3$ in the UV to the relativistic value $z=1$ in the 
IR\@.  In this paper, we present some evidence suggesting that the CDT 
approach to lattice gravity may in fact be a lattice version of the quantum 
gravity at a Lifshitz point.  Using the same definition of dimension as in 
the CDT approach \cite{ajl}, we show that in the continuum framework of 
\cite{lif} the effective dimension of the Universe flows from four at 
large distances to two at short distances, reproducing the lattice results 
of \cite{ajl}.  

\section{The Spectral Dimension of Fluctuating Geometries}

In principle, there are many different ways of defining the dimension of a 
fluctuating geometry.  Here we follow \cite{ajl}, and consider a measure of 
dimension which has proven useful in discretized approaches to quantum 
gravity in low dimensions:  the ``spectral dimension'' of spacetime.  
The idea is simple: Spectral dimension of a geometric object $\CM$ is the 
effective dimension of $\CM$ as seen by an appropriately defined diffusion 
process (or a random walker) on $\CM$.  The diffusion process is characterized 
by the probability density $\rho(\bw,\bw';\sigma)$ of diffusion from point 
$\bw$ to point $\bw'$ in $\CM$, in diffusion time $\sigma$, subjected to the 
intial condition $\rho(\bw,\bw';0)=\delta(\bw-\bw')$.   The {\it average 
return probability\/} $P(\sigma)$ is obtained by evaluating 
$\rho(\bw,\bw';\sigma)$ at $\bw=\bw'$ and averaging over all points $\bw$ in 
$\CM$.  The {\it spectral dimension} of $\CM$ is then defined in terms of 
$P(\sigma)$,
\be
\label{specd}
d_s=-2\frac{d\,\log P(\sigma)}{d\,\log\sigma}.  
\ee
For example, in the case of $\CM=\R^d$ with the flat Euclidean metric, we 
obtain
\be
\label{denseu}
\rho(\bw,\bw';\sigma)=\frac{e^{-(\bw-\bw')^2/4\sigma}}{(4\pi\sigma)^{d/2}}.
\ee
In this case, the spectral dimension (\ref{specd}) is $d_s(\R^d)=d$, which 
simply reproduces the topological dimension of the Euclidean space. 

The spectral dimension can be defined in a manifestly coordinate-independent 
way, which makes it applicable to a wide range of geometric objects beyond 
smooth manifolds, including those with various forms of fractal behavior.  
Indeed, objects are known for which $d_s$ is not an integer:  For example, the 
spectral dimension of branched polymers \cite{ao} is $d_s=4/3$.  

The spectral dimension has been used \cite{wheater1,wheater2,wheater3,%
ambjorn1,ambjorn2,ambjorn3,durhuus1,durhuus2} as one of the simplest 
observables probing the continuum limit in the lattice approach to quantum 
gravity in two dimensions.  This case is relevant for the description of 
fluctuating worldsheets in noncritical string theory.  In the nonperturbative 
definition of the system in terms of dynamical triangulations and matrix 
models, the spectral dimension of worldsheets has been found to be $d_s=2$ 
\cite{ambjorn2}, as long as the central charge of the worldsheet matter sector 
is $c\leq 1$.  Above this $c=1$ barrier, the ensemble of fluctuating 
geometries is believed to collapse to a branched polymer 
phase.  This expectation has been further confirmed by the measurement of the 
spectral dimension in \cite{wheater1}, yielding $d_s=4/3$ above $c=1$.  
Interestingly, this simplest branched polymer phase of two-dimensional gravity 
is in fact the lowest member of an infinite family of multi-critical phases 
\cite{wheater2}, parametrized by $m=2,3,\ldots$, and with spectral dimensions 
\be
\label{multibp}
d_s=2m/(2m-1).  
\ee

In \cite{ajl}, the spectral dimension of spacetime was measured in the 
numerical CDT approach to quantum gravity in $3+1$ dimensions, with intriguing 
results.  At long distances, the spectral dimension found by \cite{ajl} is
\be
\label{sdfour}
d_s=4.02\pm 0.1,
\ee
\ie , the spacetime is macroscopically four-dimensional.  With the changing 
scale, however, the spectral dimension appears to smooothly decrease to the 
short-distance limit, given by \cite{ajl}
\be
\label{sdtwo}
d_s=1.80\pm 0.25.
\ee
This value is consistent with the effective reduction of spacetime to two 
dimensions at short distances.  

As we will show, a similar reduction in the spectral dimension of spacetime 
is found in the continuum path-integral approach to quantum gravity with 
anisotropic scaling, presented in \cite{mqc,lif}.  

\section{Gravity at a Lifshitz Point}

The anisotropic scaling of spacetime is characterized by the dynamical 
critical exponent $z$,
\be
\label{scaling}
\bx\to b\bx,\qquad t\to b^zt.
\ee
Models with anisotropic scaling are common in condensed matter (see, \eg , 
\cite{sachdev}).  Theories of gravity with various values of $z$ in 
various spacetime dimensions $D+1$ were introduced in \cite{mqc,lif}.  The 
case of Yang-Mills with $z=2$ was discussed in \cite{cym}.

For power-counting renormalizability of gravity in $3+1$ dimensions, we need 
$z=3$ at short distances \cite{lif} (see also \cite{visser}).  A theory of 
gravity in $3+1$ dimensions with $z=3$ was presented in \cite{lif}.  The field 
content consists of the spatial metric $g_{ij}$, together with the lapse and 
shift variables $N_i$ and $N$.  The theory is invariant under 
foliation-preserving diffeomorphisms $\dif(M)$ of spacetime, which take the 
coordinate form
\be
\label{fdiff}
\tilde x^i=\tilde x^i(t,x^j),\qquad \tilde t=\tilde t (t).
\ee
The action is given by
\be
\label{lifact}
S=\frac{2}{\kappa^2}\int dt\,d^3\bx\,\sqrt{g}\,N\left\{K_{ij}K^{ij}-\lambda
\left(K_i^i\right)^2-\CV\right\}.
\ee
Here 
\be
K_{ij}=\frac{1}{2N}\left(\dot g_{ij}-\nabla_iN_j-\nabla_jN_i\right)
\ee
is the extrinsic curvature tensor of the preferred time foliation $\CF$ of 
spacetime.  
In gravity with anisotropic scaling and $\dif(M)$ gauge symmetry, 
$K_{ij}$ plays the role of the covariant time derivative of the metric 
tensor.  
The first two terms in (\ref{lifact}) represent the covariant kinetic 
term, of second order in the time derivatives of the metric, with $\kappa$ 
and $\lambda$ two dimensionless couplings left undetermined by the gauge 
symmetries of $\dif(M)$.  

The potential term $\CV$ in (\ref{lifact}) is a local function of 
$g_{ij}$ and its spatial derivatives, but independent of $\dot g_{ij}$.  
Unlike the kinetic term quadratic in $K_{ij}$, which is universal 
and independent of the choice of $z$, the precise form of $\CV$ depends on the 
desired value of $z$.  For example, general relativity requires 
$\CV\propto R-2\Lambda$ (and $\lambda=1$, to satisfy full spacetime 
diffeomorphism invariance), implying of course the relativistic value of 
$z=1$.   

In condensed matter, a particularly interesting class of models with $z\neq 1$ 
satisfies an additional condition of ``detailed balance.''  Those models 
are intimately related to a Euclidean theory in one lower dimension.  In the 
case of gravity in $3+1$ dimensions, this condition means that 
$\CV\sim(\delta W/\delta g_{ij})^2$, where $W$ is the action of a gravity 
theory in three dimensions.  (The square is performed with the appropriate 
De~Witt metric; see \cite{mqc,lif} for details).  

The $z=3$ gravity introduced in \cite{lif} is described by (\ref{lifact}) with 
\be
\label{vcc}
\CV=\frac{\kappa^4}{16w^4}C_{ij}\,C^{ij},
\ee
where $C_{ij}$ is the Cotton tensor,
\be
\label{cotton}
C^{ij}=\epsilon^{ik\ell}\nabla_k\left(R_\ell^j\right.-\frac{1}{4}\left.
\delta_\ell^jR\right),
\ee
and $w$ is a dimensionless coupling constant.  Since $C_{ij}=0$ is the 
equation of motion following from the variation of the Chern-Simons action 
$W=(1/w^2)\int \left(\Gamma\wedge d\Gamma+\frac{2}{3}\Gamma\wedge\Gamma\wedge
\Gamma\right)$, the theory satisfies detailed balance.  This condition can 
be explicitly broken by the addition of diffeomorphism invariant terms of 
sixth order in spatial derivatives, such as $\nabla_iR_{jk}\nabla^iR^{jk}$ or 
$R^i_jR^j_kR^k_i$, to the potential $\CV$.  This does not change the value of 
$z$, but theories without detailed balance are generally more complex due to
the proliferation of independent terms in the action.  Luckily, the spectral 
dimension that we consider below turns out to be a very universal observable, 
sensitive only to the scaling (\ref{scaling}) but not to the details of 
$\CV$.  

The global scaling transformations (\ref{scaling}) can be generalized to the 
case when the spacetime geometry is fluctuating and the background is no 
longer flat.  In \cite{lif}, the local anisotropic Weyl tranformations with 
$z=3$ were introduced,  
\be
\label{anweyl}
g_{ij}\to e^{2\Omega(\bx,t)}g_{ij},\qquad
N_i\to e^{2\Omega(\bx,t)}N_i,\qquad
N\to e^{3\Omega(\bx,t)}N.
\ee
(Other values of $z$ were discussed in \cite{mqc}.)  These represent a local 
version of the global anisotropic scaling (\ref{scaling}) of flat space, 
adapted to the general background $g_{ij}$, $N_i$ and $N$.  The anisotropic 
Weyl transformations (\ref{anweyl}) form a closed symmetry group with the 
foliation-preserving diffeomorphisms $\dif(M)$ of (\ref{fdiff}) 
(see \cite{mqc,lif}).  Since the Cotton tensor transforms covariantly under 
local conformal transformations of space, the potential term (\ref{vcc}) is 
invariant under (\ref{anweyl}).  At the special value of $\lambda=1/3$, the 
kinetic term also becomes invariant under (\ref{anweyl}).

In $z=3$ gravity, the leading $C^2$ term (\ref{vcc}) in $\CV$ is of the same 
dimension as the kinetic term $\sim K^2$, and dominates the potential at 
short distances.  However, the gauge symmetries of $\dif(M)$ allow a 
number of relevant terms in $\CV$, which affect the dynamics at long 
distances.  The theory flows in the infrared to lower values of $z$, and 
ultimately to $z=1$.  Such relevant terms in $\CV$ can be generated without 
violating detailed balance by adding two relevant terms to the 
three-dimensional Chern-Simons action $W$: the Ricci scalar $R$ and the 
cosmological constant term.  This turns $W$ into the action of topologically 
massive gravity, and results in the modified potential
\be
\label{defv}
\CV=\frac{\kappa^4}{16w^4}C_{ij}\,C^{ij}+\ldots-\frac{c^2}{2\kappa^2}
(R-2\Lambda).
\ee
(The ``$\ldots$'' in (\ref{defv}) stand for intermediate terms of fourth and 
fifth order in spatial derivatives.)

From the perspective of the $z=3$ UV fixed point, $c$ and $\Lambda$ are 
relevant coupling constants, of dimension two (in the units of inverse spatial 
length).  The last two terms in (\ref{defv}) are those 
that appear in the potential $\CV$ of general relativity.  At long distances, 
it is natural to redefine the time coordinate to reflect the $z=1$ scaling, by 
setting $x^0=ct$. The theory in the infrared then closely resembles general 
relativity, with the effective Newton constant given by 
$\newton=\kappa^2/(32\pi c)$.

\section{The Spectral Dimension of Spacetimes with Anisotropic Scaling}

In order to compare the behavior of the spectral dimension in the 
lattice CDT approach \cite{ajl} with the analytic approach of gravity at a 
Lifshitz point, we must extend the definition of spectral dimension to smooth 
spacetimes with anisotropic scaling (\ref{scaling}).  

What is the appropriate diffusion process to consider?  Recall first 
\cite{ajl} that in the relativistic case of $z=1$, the spectral dimension of 
the Minkowski spacetime is measured by first rotating to imaginary time, 
$t=-i\tau$.  On the resulting Euclidean space, the diffusion process is 
described by the probability density $\rho$ of (\ref{denseu}), governed by the 
diffusion equation
\be
\frac{\p}{\p\sigma}\rho(\bx,\tau;\bx',\tau';\sigma)=\left(\frac{\p^2}{\p\tau^2}
+\Delta\right)\rho(\bx,\tau;\bx',\tau';\sigma),
\ee
where $\Delta\equiv\p_i\p_i$ is the {\it spatial\/} Laplacian.  This can be 
naturally generalized to the case with dynamical critical exponent $z\neq 1$.  
In the theories of gravity with anisotropic scaling, the dynamics along the 
time dimension stays qualitatively the same as in the $z=1$ case (\ie , the 
Lagrangian is of the same order in time derivatives).  It is the spatial 
dynamics that changes with the changing potential $\CV$.  This suggests that 
the natural diffusion process at general $z$ is governed by the anisotropic 
diffusion equation,
\be
\label{andif}
\frac{\p}{\p\sigma}\rho(\bx,\tau;\bx',\tau';\sigma)=\left\{
\frac{\p^2}{\p\tau^2}+(-1)^{z+1}\Delta^z\right\}\rho(\bx,\tau;\bx',\tau;
\sigma).
\ee
Indeed, both terms in the operator on the right-hand-side of the equation 
scale the same way under the anisotropic rescaling of space and time 
(\ref{scaling}).  The relative sign $(-1)^{z+1}$ in (\ref{andif}) is 
determined from the requirement of ellipticity of the diffusion operator.  
The formula is valid for integer values of $z$, but our results below can be 
analytically continued to any positive real value 
of $z$.  

The anisotropic diffusion equation (\ref{andif}) is solved by
\be
\rho(\bx,\tau;\bx',\tau';\sigma)=\int\frac{d\omega\,d^D\bk}{(2\pi)^{D+1}}
e^{i\omega (\tau-\tau')+i\bx\cdot(\bx-\bx')}
e^{-\sigma(\omega^2+|\bk|^{2z})}.
\ee
In order to determine the spectral dimension, we only need $\rho$ at the 
coincident initial and final spacetime points, $\bx=\bx'$ and $\tau=\tau'$, 
\be
\rho(\bx,\tau;\bx,\tau;\sigma)=\int\frac{d\omega\,d^D\bk}{(2\pi)^{D+1}}
e^{\sigma(\omega^2+|\bk|^{2z})}=
\frac{C}{\sigma^{(1+D/z)/2}},
\ee
with some nozero constant $C$.  Using (\ref{specd}), we finally obtain the 
spectral dimension of spacetime with anisotropic scaling,
\be
\label{dsz}
d_s\equiv-2\frac{d\,\log P(\sigma)}{d\,\log\sigma}=1+\frac{D}{z}.
\ee
This implies the central result of this paper:  In the $3+1$ dimensional 
spacetime with $z=3$, the spectral dimension (\ref{dsz}) is equal to $d_s=2$.  
Under the influence of the relevant deformations, the theory flows 
to $z=1$ in the infrared, reproducing the macroscopic value $d_s=4$ at long 
distances.    

This result has been evaluated for a fixed classical spacetime geometry, 
described by 
\be
\label{flatst}
g_{ij}=\delta_{ij},\qquad N=1,\qquad N_i=0.  
\ee
Consequently, (\ref{dsz}) represents the leading value for $d_s$ in the 
semiclassical approximation.   The notion of the spectral dimension can be 
generalized to the full quantum path integral of the system, by 
defining the covariant generalization of the diffusion operator on an 
arbitrary curved geometry, and averaging the return probability over all 
configurations in the path integral.  For theories which exhibit the flat 
spacetime (\ref{flatst}) as a classical solution, (\ref{dsz}) represents the 
leading term in the semiclassical evaluation of the spectral dimension.  
With quantum corrections small at large distances, the dependence of the 
spectral dimension on the spacetime scale is dominated by the change in the 
anisotropic scaling of the classical solution, from $z=3$ in the UV to $z=1$ 
in the IR.  

An effective cosmological constant in the IR might preclude the flat 
geometry (\ref{flatst}) from being a classical solution.  If that happens, 
the spectral dimension will become sensitive at cosmological scales to the 
characteristic curvature of spacetime.  However, such finite-size effects will 
not change the effective spectral dimension (\ref{dsz}) at intermediate 
scales. 

\section{Discussion}

We have extended the notion of spectral dimension to the continuum framework 
of quantum gravity with anisotropic scaling, and found that the behavior 
of spectral dimension matches qualitatively the lattice results obtained by 
Ambj\o rn {\it et al} \cite{ajl} in the CDT approach.  This raises the 
intriguing possibility that the continuum limit of the causal dynamical 
triangulations may belong to the same universality class as the anisotropic 
theory of gravity \cite{lif}, flowing from the anisotropic scaling with $z=3$ 
in the UV to the relativistic value $z=1$ in the IR.  

This possibility that the CDT lattice approach is effectively a lattice 
description of quantum field theory of gravity with anisotropic scaling 
presented in \cite{mqc,lif} is further supported by the symmetries imposed in 
the two frameworks.   As reviewed above, theories of gravity with anisotropic 
scaling are invariant under foliation-preserving diffeomorphisms; the 
spacetime manifold is equipped with a preferred causal structure, compatible 
with anisotropic scaling (see Fig~1a).  
On the other hand, the novelty of the CDT approach to lattice gravity is that 
the sum is performed over lattice geometries with a preferred ``causal 
structure'' (Fig.~1b).  Indeed, it is this extra condition on the 
discretizations which changes favorably the continuum limit, and prevents 
the collapse of the partition sum to a branched polymer phase.  It is 
plausible that the continuum limit of the lattice sum automatically identifies 
a mechanism leading to its UV completion in the minimal way compatible with 
the preferred foliation, \ie , in terms of a gravity theory with anisotropic 
scaling and $z=3$ at short distances.  
\EPSFIGURE{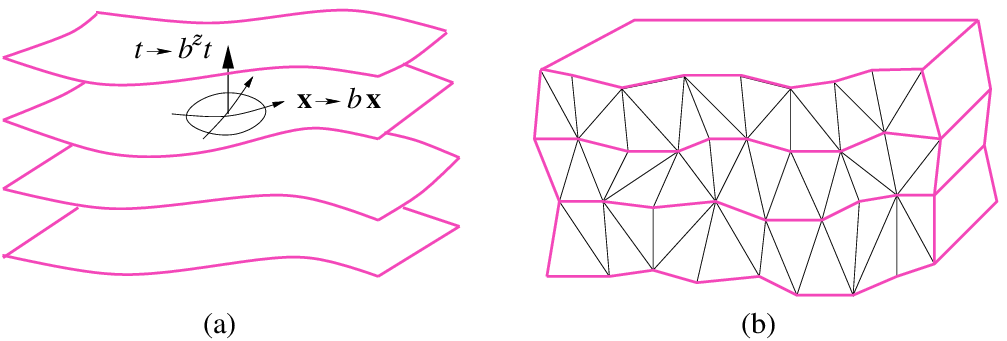}{
(a) The preferred foliation by time slices in the continuum approach of 
gravity with anisotropic scaling; (b) a characteristic configuration in 
the causal dynamical triangulation approach.}  

The short-distance lattice value (\ref{sdtwo}) of the spectral dimension 
is consistent within the margin of error with $z=3$.  However, the mean value 
reported is in fact closer to the spectral dimension of a smooth spacetime 
with $z=4$.  Theories with $z=4$ in $3+1$ dimensions satisfying the detailed 
balance condition were discussed in \cite{lif}:  They are constructed from the 
three-dimensional action $W$ containing terms up to quadratic in the Ricci 
tensor.  (Such models of three-dimensional gravity have recently been 
discussed in \cite{bt}.)  Reasons why gravity with $z=4$ might be desirable 
in $3+1$ dimensions were discussed in \cite{lif}.   

Even though the main focus of this paper is on gravity in $3+1$ dimensions, 
our result (\ref{dsz}) for the spectral dimension of spacetime with 
anisotropic scaling is general, with possible applications to quantum 
gravity in other dimensions.  For example, it is intriguing that the spectral 
dimensions (\ref{multibp}) observed in the multicritical branched-polymer 
phases of discretized two-dimensional gravity can be reproduced by continuum 
theories in $1+1$ dimensions with anisotropic scaling and the integer 
multicritical values of the dynamical exponent $z=2m-1$.  

The spectral dimension also plays a prominent role in the thermal behavior 
of systems with anisotropic scaling.  For example, simple scaling arguments 
show that the free energy of a system of free massless fields at the Lifshitz 
point with dynamical critical exponent $z$ scales with temperature as 
$F\sim T^{1+D/z}=T^{d_s}$.  Notably, when $D=z$ (the critical dimension of 
gravity with anisotropic scaling), the behavior of the free energy $F\sim T^2$ 
is the same as in relativistic CFT in $1+1$ dimensions.  This scaling 
has been seen before, by Atick and Witten \cite{atick} in their study of the 
ensemble of free strings, formally extrapolated into the regime above the 
Hagedorn temperature.  An example of anisotropic gravity with $z=9$ in $9+1$ 
dimensions can be obtained by following the logic of \cite{lif}:  Starting 
with $W\sim\int\omega_9$, with $\omega_9=\Gamma\wedge (d\Gamma)^4+\ldots$ the 
Chern-Simons 9-form, and setting $\CV=(\delta W/\delta g_{ij})^2$ leads to a 
theory with detailed balance in $9+1$ dimensions with $z=9$, whose 
high-temperature behavior at the free-field fixed point matches the scaling 
found in \cite{atick} in ten-dimensional superstring theory formally 
extrapolated above the Hagedorn temperature.  

In conclusion: We have demonstrated that even for smooth spacetime geometries, 
the spectral dimension of spacetime does not have to match its topological 
dimension.  The discrepancy between the two can simply result from anisotropic 
scaling, compatible with a preferred causal structure of spacetime.  This 
suggests an alternative interpretation of the dynamical reduction of 
spacetime at short distances \cite{ajl} observed in the lattice aproach to 
quantum gravity:  This behavior does not have to indicate a change in the 
topological dimension of spacetime, or a foamy structure in which the four 
macroscopic dimensions result from coarse graining over topologically 
complicated two-dimensional geometries.  Instead, the 
observed behavior of \cite{ajl} can simply be explained by anisotropic 
scaling of space and time at short distances, keeping the topology of 
spacetime four-dimensional and its geometry smooth and topologically trivial.  

\acknowledgments

This work has been supported in part by NSF Grant PHY-0555662, DOE Grant 
DE-AC03-76SF00098, and the Berkeley Center for Theoretical Physics.  

\bibliographystyle{JHEP}
\bibliography{spe}
\end{document}